\documentclass[english]{article}
\usepackage[T1]{fontenc}
\usepackage[latin9]{inputenc}
\usepackage{amsthm}
\usepackage{amsmath}
\usepackage{amssymb}
\usepackage{graphicx}
\usepackage{epstopdf}
\usepackage{spconf}
\allowdisplaybreaks

\makeatletter
\theoremstyle{plain}

  \theoremstyle{remark}

\name{Dionysios S. Kalogerias$^{\dagger}$,
	  Nikolaos Chatzipanagiotis$^{\star}$, 
	  Michael M. Zavlanos$^{\star}$ and 
	  Athina P. Petropulu$^{\dagger}$
\thanks{This work is supported by the National Science Foundation under Grant CNS-09-05425.}}

\address{$^{\dagger}$ Department of Electrical and Computer Engineering,\\
                  	  Rutgers, The State University of New Jersey, 
                  	  New Brunswick, NJ 08854, USA \\
         $^{\star}$ Department of Mechanical Engineering and Materials Science,\\
                    Duke University, Durham, NC 27708, USA}

\makeatother

\usepackage{babel}
  \providecommand{\remarkname}{Remark}
\providecommand{\theoremname}{Theorem}

\begin{document}

\title{Mobile Jammers for Secrecy Rate Maximization
in\\ Cooperative Networks}
\maketitle
\begin{abstract}

%\footnote{Work supported by the National Science Foundation under Grant
%CNS-09-05425} %We address the problem of improving physical layer security in wireless
%communication networks abetted by mobile  helpers. 
We consider
a source (Alice) trying to communicate with a destination (Bob), in a way that an unauthorized node (Eve)
cannot infer, based on her observations, the information that is being transmitted. 
The communication is assisted by multiple multi-antenna cooperating nodes (helpers) who have the ability to move.
 While Alice transmits, the helpers transmit noise that is designed to affect the entire space except Bob.
We consider the problem of selecting the helper weights and positions that maximize the 
system secrecy rate. It turns out that this optimization problem can be efficiently
solved, leading to a novel decentralized helper motion control scheme.
Simulations indicate that introducing helper mobility leads to considerable savings in terms of helper transmit power,
as well as  total number of helpers required for secrecy communications.
\end{abstract}

\section{Introduction}

Information secrecy is a challenge in wireless communications, as the wireless channel makes the transmitted information accessible to  unauthorized   as well as  legitimate nodes. Physical (PHY) layer secrecy  exploits channel conditions to maximize the rate at which reliable information is delivered to the legitimate destination, with unauthorized users having the maximum possible uncertainty about the transmitted signal based on their observations
 \cite{wyner}. 
 %it was shown  that when the source-wiretapper channel is weaker that the source-destination channel, the source and the destination can exchange messages at a non-zero rate in perfect secrecy, i.e., the wiretapper can learn almost nothing about the messages based on its observations.
In wireless communications, secrecy is typically addressed with cryptographic approaches, which rely on secret keys that are distributed to the network users periodically. However, breaking a secret key is only a matter of computational power and time. Although longer keys which are updated often are more difficult to be broken, however, they occupy valuable communication bandwidth. On the other hand, physical layer secrecy,
 exploits the randomness of the wireless channel to maximize the ambiguity of the transmitted signal at an unauthorized receiver. Secrecy communication through broadcast channels was studied in \cite{Csiszar} and for the scalar Gaussian wiretap channel in \cite{Hellman}.
While for  single-input single-output (SISO) channels perfect secrecy is achievable only when the  Alice-Eve channel is a degraded version of the Alice-Bob channel \cite{wyner, Hellman},
the use of multiple antennas \cite{Khisti2, Swindlehurst3} can ensure positive secrecy rate even when the SISO methods fail.
Cooperative jamming (CJ)  is another way to achieve high secrecy rates, and can  be used when the wireless transceiver, due to size limitations, cannot accommodate more than one antennas. In CJ, helpers transmit noise to degrade the Alice-Eve channel. In most existing CJ schemes the helpers require full channel state information (CSI) on Eve \cite{Dong10TSP,Gan}, which implicitly assumes that Eve is a known node in the network. Works that approach the problem requiring partial CSI on Eve, or no CSI at all, include \cite{Swindlehurst,Luo}.

Mobility has been shown to dramatically improve the capacity of multiuser ad  hoc wireless networks  with random relay-assisted source-destination pairs \cite{TSE1}.
In \cite{TRAP1, TRAP2}, distributed network mobility control is employed in order to guide the network nodes for achieving specific operational goals, such as the adaptation of the network topology to a jamming attack,  motivated by classical mechanics and through the use of carefully designed sequential algorithms. Also, more recently, in \cite{ChHawaii2012}, mobility control has been combined with optimal beamforming for transmit power minimization under Quality of Service (QoS) constraints in multiuser cooperative networks.

This paper follows up on the recent work by some of the authors \cite{Luo},
in which, the communication between Alice and Bob is aided by multi-antenna helpers.
Each helper transmits noise that is in the null-space of the helper's channel to Bob, thus creating interferences at all points in space, which include Eve, but not Bob.
No  coordination between helpers is required to design the noise, and no  CSI on Alice is needed.  By imposing nulling at Bob, the scheme of \cite{Luo} is suboptimal, however, it was shown \cite{Luo} to perform very close to a scheme that transmits optimally designed noise.
By exploiting  
the dependence of jamming noise on the helper positions, we propose a method that further improves the secrecy rate performance. 
%We propose a hybrid control
%scheme, where the jamming noise design is integrated with potential-field-based
%motion control, designed to maximize the secrecy rate subject to power constraints.
In particular, we consider the problem of jointly maximizing the system secrecy
rate with respect to the helper weights and  positions. It turns out that this joint optimization problem can be efficiently
solved, leading to a novel decentralized helper motion control scheme.
Numerical simulations show that the proposed scheme can
lead to considerable savings in terms of helper transmitting power,
as well as the total number of helpers required for meeting a certain secrecy rate.

Our work assumes that each helper has a map of its channel to Bob, Alice and Eve.
Such map can be constructed based on the exchange of messages, which assumes that Eve is a known quantity in the network; for example, Eve is not an enemy, but rather a node that has limited access to information and certain communications need to be kept secret from her.

The rest of this paper is organized as follows. In Section 2, we present the system model.
In Section 3, we briefly describe the nulling noise technique for stationary helpers, and in Section 4, we combine this technique with
helper mobility control, leading to our proposed decentralized motion control scheme for joint maximization of the system secrecy rate.
Finally, in Section 5, we present numerical simulations,  illustrating the effectiveness of the proposed approach.

\section{System Model}

\begin{figure}
\centering\includegraphics[scale=0.60]{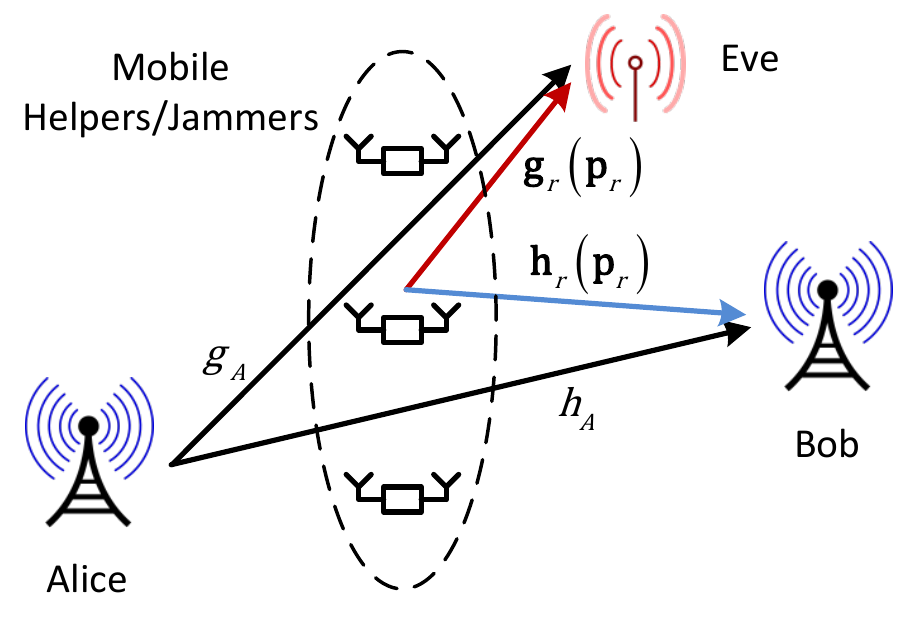}

\caption{\label{fig:System-Model}System Model}
\end{figure}

The system model is depicted in  Fig.
\ref{fig:System-Model}. Alice wishes 
to communicate with Bob, keeping the transmitted information secret from Eve.
Alice, Bob and Eve are
each equipped  with a single transmission/reception antenna and are located at positions $\mathbf{p}_{i}\in\mathbb{R}^{2},i\in\left\{ A,B,E\right\}$, respectively.

The transmission
is assisted by $R$ helpers, each of which is equipped with $N_{r}$
antennas, with $N_{r}\geq2,r\in\mathbb{N}_{R}^{+}\equiv\left\lbrace 1,2,...,R\right\rbrace $. The
area spanned by each helper is considered a disc with diameter $\rho$ and center located at position $\mathbf{p}_{r}\in\mathbb{R}^{2}$. Without any ambiguity, we will refer to $\mathbf{p}_{r}$ as the position of helper $r$.
Moreover, the antennas of each helper $r$ are located at positions $\left(\mathbf{p}_{r} + \mathbf{p}_{r}^{\rho k} \right) \in\mathbb{R}^{2},k\in\mathbb{N}_{N_r}^{+}$, so that the spacing between any two antennas on the same helper is at least $\lambda/2$, where $\lambda$ denotes the carrier wavelength used for the communication.

We assume that, throughout each transmission cycle, Alice,
Bob and Eve are stationary, while the helpers can independently move at
will. Based on this assumption, the channels from Alice to Bob and
Eve are denoted by $h_{A}\left(\mathbf{p}_{A},\mathbf{p}_{B}\right)\equiv h_{A}\in\mathbb{C}$
and $g_{A}\left(\mathbf{p}_{A},\mathbf{p}_{E}\right)\equiv g_{A}\in\mathbb{C}$,
respectively.
Also, the channels from each helper to Bob and Eve are
denoted by $\mathbf{h}_{r}\left(\mathbf{p}_{B},\mathbf{p}_{r}\right)\equiv\mathbf{h}_{r}\left(\mathbf{p}_{r}\right)\in\mathbb{C}^{N_{r}}$
and $\mathbf{g}_{r}\left(\mathbf{p}_{E},\mathbf{p}_{r}\right)\equiv\mathbf{g}_{r}\left(\mathbf{p}_{r}\right)\in\mathbb{C}^{N_{r}}$,
$r\in\mathbb{N}_{R}^{+}$, respectively. Assuming that rotational motion is not allowed, then, for each helper, the channel gain for each
specific antenna can be parametrized with respect to its position,
since, if $\mathbf{p}_{r}$ is determined,  the positions of all
its antennas can be deterministically set from its respective antenna
topology. More specifically, we define
\begin{flalign}
\mathbf{h}_{r}\left(\mathbf{p}_{r}\right) & \triangleq\left[h_{r1}\left(\mathbf{p}_{r}+\mathbf{p}_{r}^{\rho 1}\right)\,\ldots\, h_{rN_{r}}\left(\mathbf{p}_{r}+\mathbf{p}_{r}^{\rho N_{r}}\right)\right]^{T},\\
\mathbf{g}_{r}\left(\mathbf{p}_{r}\right) & \triangleq\left[g_{r1}\left(\mathbf{p}_{r}+\mathbf{p}_{r}^{\rho 1}\right)\,\ldots\, g_{rN_{r}}\left(\mathbf{p}_{r}+\mathbf{p}_{r}^{\rho N_{r}}\right)\right]^{T}.
\end{flalign}
%where $\mathbf{p}_{r}^{k},k\in\mathbb{N}_{N_{r}}^{+}\equiv\left\lbrace %1,2,...,N_{r}\right\rbrace$ denotes
%the position of the antenna $k$ of the helper $r$ with respect to
%the position of the helper $r$, $\mathbf{p}_{r}$ (its disc center).

Alice transmits the signal $\sqrt{P_{s}}x$, where $P_{s}$ denotes
the transmission power and $x$ is assumed to be an
arbitrary zero mean complex random variable with $\mathbb{E}\{\left|x\right|^{2}\}=1$.
Additionally, we assign an individual power budget to each helper denoted
as $P_{r},r\in\mathbb{N}_{R}^{+}$.

We consider a mobility - enabled cooperative jamming
scenario, that is, while Alice is transmitting, all helpers can independently
move and independently transmit the artificial noise signal $\mathbf{n}_{r}\left(\mathbf{p}_{r}\right)\in\mathbb{C}^{N_{r}},r\in\mathbb{N}_{R}^{+}$,
which is also independent of the information transmitted. Under this setting,
Bob and Eve receive the signals
\begin{flalign}
y_{B} & \triangleq\sqrt{P_{s}}h_{A}x+\sum_{r=1}^{R}\mathbf{h}_{r}^{T}\left(\mathbf{p}_{r}\right)\mathbf{n}_{r}\left(\mathbf{p}_{r}\right)+n_{B},\label{eq:1}\\
y_{E} & \triangleq\sqrt{P_{s}}g_{A}x+\sum_{r=1}^{R}\mathbf{g}_{r}^{T}\left(\mathbf{p}_{r}\right)\mathbf{n}_{r}\left(\mathbf{p}_{r}\right)+n_{E},\label{eq:2}
\end{flalign}
respectively, where $n_{B}$ and $n_{E}$ denote complex Additive
White Gaussian Noise (AWGN) quantities at the respective reception
points with equal variances $\mathbb{E}\{\left|n_{B}\right|^{2}\}=\mathbb{E}\{\left|n_{E}\right|^{2}\}=N_{0}$.

In the following, we assume that the quantities $\sqrt{P_s}$, $h_A$, $g_A$, $\mathbf{h}_r(\mathbf{p}_r)$, $\mathbf{g}_r(\mathbf{p}_r)$ are known.
For each position of the helpers in the environment, our goal is to design the jamming noise $\mathbf{n}_r(\mathbf{p}_r)$  and
at the same time control the positions of the helpers $\mathbf{p}_r$, $r\in\mathbb{N}_{R}^{+}$ so that the secrecy
rate of the whole system is maximized.

The wireless channel is assumed to be flat
fading. Assuming a rich scattering environment \cite{ChHawaii2012},
the baseband equivalent channel gain $c_{ij}\left(\mathbf{p}_{i},\mathbf{p}_{j}\right)\in\mathbb{C}$
between two arbitrary antennas $i$ and $j$ with respective positions
$\mathbf{p}_{i}$ and $\mathbf{p}_{j}$ and Euclidean distance $d_{ij}\left(\mathbf{p}_{i},\mathbf{p}_{j}\right)\equiv d_{ij}$,
can be approximated as $
c_{ij}\left(\mathbf{p}_{i},\mathbf{p}_{j}\right)=\alpha_{ij}\left(\mathbf{p}_{i},\mathbf{p}_{j}\right)\beta_{ij}e^{j\left(2\pi/\lambda\right)d_{ij}},
$
where $\alpha_{ij}\left(\mathbf{p}_{i},\mathbf{p}_{j}\right)\sim\mathcal{CN}\left(0,1/2\right)$
 models multipath fading, and $\beta_{ij}=d_{ij}^{-\mu/2}$ models path
loss, where $\mu$ denotes the path loss coefficient and represents
the power fall - off rate of the channel.
Further, we assume that $\alpha_{ij}\left(\mathbf{p}_{i},\mathbf{p}_{j}\right)$
and $\alpha_{kj}\left(\mathbf{p}_{k},\mathbf{p}_{j}\right)$ are independent
if $d_{ik}\geq\lambda/2$ and correlated otherwise. Thus, for a fixed
position of, for instance, an antenna $j$, we can create a spatial
map that returns the multipath fading coefficient $\alpha_{ij}\left(\mathbf{p}_{i},\mathbf{p}_{j}\right)$
(and of course $c_{ij}\left(\mathbf{p}_{i},\mathbf{p}_{j}\right)$)
for any position of the antenna $i$ in the environment.

\section{Generation of Nulling Noise}

Each helper generates Nulling Noise (NN),  so that Bob is interference - free \cite{Luo}. More specifically,
for fixed positions throughout the plane, the noise transmitted by
each helper is designed so that
\begin{equation}
\mathbf{h}_{r}^{T}\left(\mathbf{p}_{r}\right)\mathbf{n}_{r}\left(\mathbf{p}_{r}\right)=0,\quad r\in\mathbb{N}_{R}^{+}.\label{eq:3}
\end{equation}
For $N_{r}\geq2,r\in\mathbb{N}_{R}^{+}$, \eqref{eq:3} enjoys the
simple closed form solution
$\mathbf{n}_{r}\left(\mathbf{p}_{r}\right)=\mathbf{E}_{r}\left(\mathbf{p}_{r}\right)\mathbf{v}_{r}
$%\end{equation}
, where $\mathbf{E}_{r}\left(\mathbf{p}_{r}\right)\in\mathbb{C}^{N_{r}\times\left(N_{r}-1\right)}$
denotes a column - orthonormal basis matrix for the right nullspace of $\mathbf{h}_{r}^{T}\left(\mathbf{p}_{r}\right)$
and $\mathbf{v}_{r}\in\mathbb{C}^{\left(N_{r}-1\right)\times1}$ denotes
an arbitrary vector. Of course, $\mathbf{E}_{r}\left(\mathbf{p}_{r}\right)$
can be easily obtained from $\mathbf{h}_{r}^{T}\left(\mathbf{p}_{r}\right)$
through a usual singular value decomposition. By setting $\mathbf{v}_{r}\triangleq w_{r}\mathbf{t}_{r}$,
where $w_{r}\in\mathbb{C}$ and $\mathbf{t}_{r}\sim\mathcal{CN}\left(0,\mathbf{I}_{\left(N_{r}-1\right)\times\left(N_{r}-1\right)}\right)$,
we can design the jamming noise to be of the form \cite{Luo}
\begin{equation}
\mathbf{n}_{r}\left(\mathbf{p}_{r}\right)=w_{r}\mathbf{E}_{r}\left(\mathbf{p}_{r}\right)\mathbf{t}_{r}.
\end{equation}
Then, the power of the jamming noise is given by
\begin{alignat}{1}
&\mathbb{E}\left\{ \left\Vert \mathbf{n}_{r}\left(\mathbf{p}_{r}\right)\right\Vert ^{2}\right\}  =\mathbb{E}\left\{ \mathrm{Tr}\left(\mathbf{n}_{r}\left(\mathbf{p}_{r}\right)\mathbf{n}_{r}^{H}\left(\mathbf{p}_{r}\right)\right)\right\} \nonumber \\
 & =\left|w_{r}\right|^{2}\mathrm{Tr}\left(\mathbf{E}_{r}\left(\mathbf{p}_{r}\right)\mathbf{E}_{r}^{H}\left(\mathbf{p}_{r}\right)\right)\nonumber =\left(N_{r}-1\right)\left|w_{r}\right|^{2}\leq P_{r}.%,\quad r\in\mathbb{N}_{R}^{+}.
\end{alignat}
Note that the noise power is independent of the position of each helper.

In Section 4, we combine the NN technique with helper mobility control
so that the secrecy rate of the system defined in Section 2 is maximized.

\section{Joint Decentralized Mobility Control and  Nulling Noise for Secrecy
Rate Maximization}

Using the NN approach, the secrecy rate of the system can be expressed as
\cite{Luo}
\begin{multline}
R\left(\mathbf{w},\mathbf{p}\right)=\mathrm{log}_{2}\left(1+\dfrac{P_{s}\left|h_{A}\right|^{2}}{N_{0}}\right)\\
-\mathrm{log}_{2}\left(1+\dfrac{P_{s}\left|g_{A}\right|^{2}}{{\displaystyle \sum_{r=1}^{R}}\left|w_{r}\right|^{2}\phi_{r}\left(\mathbf{p}_{r}\right)+N_{0}}\right),\label{eq:12}
\end{multline}
where
%
%$\mathbf{w}=
%\left[\left|w_{1}\right|^{2},\ldots,\left|w_{R}\right|^{2}\right]^{T}\in\mathbb{R}^{R}, \\
% \mathbf{p}=
%\left[\mathbf{p}_{1}^{T},\ldots,\mathbf{p}_{R}^{T}\right]^{T}\in\mathbb{R}^{2R}$
%\begin{flalign}
%\mathbf{w}\triangleq
%%&
%\left[\left|w_{1}\right|^{2},\ldots,\left|w_{R}\right|^{2}\right]^{T}\in\mathbb{R}^{R},\\
%and \\$
%\phi_{r}\left(\mathbf{p}_{r}\right)= \left\Vert \mathbf{E}_{r}^{H}\left(\mathbf{p}_{r}\right)\mathbf{g}_{r}\left(\mathbf{p}_{r}\right)\right\Vert ^{2}\in\mathbb{R}^{+},\quad %r\in\mathbb{N}_{R}^{+}.$\\
%
%\label{eq:haha}
%\end{flalign}
%
\begin{flalign}
\mathbf{w}\triangleq & \left[\left|w_{1}\right|^{2},\ldots,\left|w_{R}\right|^{2}\right]^{T}\in\mathbb{R}^{R},\\
\mathbf{p}\triangleq & \left[\mathbf{p}_{1}^{T},\ldots,\mathbf{p}_{R}^{T}\right]^{T}\in\mathbb{R}^{2R}\quad\mathrm{and}\\
\phi_{r}\left(\mathbf{p}_{r}\right)\triangleq & \left\Vert \mathbf{E}_{r}^{H}\left(\mathbf{p}_{r}\right)\mathbf{g}_{r}\left(\mathbf{p}_{r}\right)\right\Vert ^{2}\in\mathbb{R}^{+},\quad r\in\mathbb{N}_{R}^{+}.\label{eq:haha}
\end{flalign}
We are interested in the joint maximization of the system secrecy
rate with respect to the weight vector $\mathbf{w}$ and the positions
vector $\mathbf{p}$, while satisfying the power budget constraints
$P_{r},r\in\mathbb{N}_{R}^{+}$ for each helper, that is, we are interested
in the optimization problem
\begin{equation}
\begin{aligned}\underset{\mathbf{w},\mathbf{p}}{\mathrm{max}}\quad & R\left(\mathbf{w},\mathbf{p}\right)\\
\mathrm{s.t}.\quad & \left|w_{r}\right|^{2}\leq\frac{P_{r}}{N_{r}-1},\quad r\in\mathbb{N}_{R}^{+}
\end{aligned}
.\label{eq:16}
\end{equation}
Obviously, $R\left(\mathbf{w},\mathbf{p}\right)$ is bounded from above
by the first term on the right-hand side of (\ref{eq:12}), 
%\begin{equation}
%\underset{\mathbf{w},\mathbf{p}}{\mathrm{sup}}\, R\left(\mathbf{w},\mathbf{p}\right)=\mathrm{log}_{2}\left(1+\dfrac{P_{s}\left|h_{A}\right|^{2}}{N_{0}}\right)\triangleq R_{sup},
%\end{equation}
where that bound is clearly not attainable for any admissible choice
of $\mathbf{w}$ and/or $\mathbf{p}$ and thus constitutes a reference
supremum for \eqref{eq:16}.

By definition of the system secrecy rate in \eqref{eq:12}, the optimization problem \eqref{eq:16} is equivalent to
\begin{equation}
\begin{aligned}
\underset{\mathbf{w}}{\mathrm{max}} \, \underset{\mathbf{p}}{\mathrm{max}}\quad & {\displaystyle \sum_{r=1}^{R}}\left|w_{r}\right|^{2}\phi_{r}\left(\mathbf{p}_{r}\right)\\
\mathrm{s.t}.\quad & \left|w_{r}\right|^{2}\leq\dfrac{P_{r}}{N_{r}-1},\quad r\in\mathbb{N}_{R}^{+}
\end{aligned}.\label{eq:17}
\end{equation}
Obviously, at the optimal solution of \eqref{eq:17} with respect
to $\mathbf{w}$, all the constraints will be active and as a result,
\eqref{eq:17} is equivalent to
\begin{equation}
\underset{\mathbf{p}}{\mathrm{max}}\quad{\displaystyle \sum_{r=1}^{R}}\beta_{r}\phi_{r}\left(\mathbf{p}_{r}\right),\label{eq:18}
\end{equation}
which is a simple unconstrained optimization problem and where
$\beta_{r}\triangleq{P_{r}}/{N_{r}-1},\quad r\in\mathbb{N}_{R}^{+}.
$
%\end{equation}
We see that regardless of its position, helper $r$ should use its
maximum power budget $P_{r}$ for noise transmission.

Now, we will additionally assume that each helper reacts to the continuous
time control input $\mathbf{u}_{r}\in\mathbb{R}^{2}$, according to
the first order differential equation
\begin{equation}
\dot{\mathbf{p}}_{r}=\mathbf{u}_{r},\quad r\in\mathbb{N}_{R}^{+}.
\end{equation}
Under this assumption, our goal is to determine the motion controllers
$\mathbf{u}_{r},\forall\, r\in\mathbb{N}_{R}^{+}$ that ensure maximization
of the secrecy rate of the system $R\left(\mathbf{w},\mathbf{p}\right)$,
or equivalently the solution of the optimization problem \eqref{eq:18}.

Similar to \cite{ChHawaii2012}, by introducing an additional collision constraint among each helper and all the other nodes in the network,
and by noting that $\beta_{r}>0,\forall\, r\in\mathbb{N}_{R}^{+}$,
we can obtain the desired controllers $\mathbf{u}_{r},r\in\mathbb{N}_{R}^{+}$
as the negative gradient of appropriately chosen artificial potential
functions for each helper $\phi_{r}^{o}:\mathbb{R}^{2}\rightarrow\mathbb{R}^{+},r\in\mathbb{N}_{R}^{+}$,
defined as
\begin{equation}
\phi_{r}^{o}\left(\mathbf{p}_{r}\right)=\phi_{r}^{col}\left(\mathbf{p}_{r}\right)-\phi_{r}\left(\mathbf{p}_{r}\right).\label{eq:19}
\end{equation}
In \eqref{eq:19}, $\phi_{r}\left(\mathbf{p}_{r}\right)$ is given
by \eqref{eq:haha} and
\begin{equation}
\phi_{r}^{col}\left(\mathbf{p}_{r}\right)\triangleq\sum_{l\in\mathcal{S}}\dfrac{1}{\left\Vert \mathbf{p}_{r}-\mathbf{p}_{l}\right\Vert ^{2}-\rho^{2}},\label{eq:20}
\end{equation}
where $\mathcal{S}\triangleq\left\{ \mathbb{N}_{R}^{+},A,B,E\right\} -\left\{ r\right\} $.
%Note that in \eqref{eq:20} we do not include the position of Eve. In fact, it is reasonable
%to assume that Eve cannot be observed by any of helpers, because if it could, then one helper could move towards
%Eve and eventually turn her physically off. Indeed, this means that
%Eve is located in a parallel 2D plane with respect to the other objects
%in the network.
%\begin{rem}
%If we assume that there is some other dedicated mechanism for collision
%avoidance in each helper (e.g. GPS), we could indeed disregard the
%collision potential completely. In any case, we include it here for
%completeness and we further assume that it is distributed by nature.
%\end{rem}

After defining the potentials $\phi_{r}^{o}\left(\mathbf{p}_{r}\right),r\in\mathbb{N}_{R}^{+}$,
the desired helper controllers are obtained as
\begin{equation}
\mathbf{u}_{r}=-\nabla_{\mathbf{p}_{r}}\phi_{r}^{o}\left(\mathbf{p}_{r}\right),\quad\forall\, r\in\mathbb{N}_{R}^{+},
\end{equation}
leading to the set of closed loop systems
\begin{equation}
\dot{\mathbf{p}}_{r}=-\nabla_{\mathbf{p}_{r}}\phi_{r}^{o}\left(\mathbf{p}_{r}\right),\quad r\in\mathbb{N}_{R}^{+}.\label{eq:23}
\end{equation}

It is evident that the gradient $\nabla_{\mathbf{p}_{r}}\phi_{r}^{o}\left(\mathbf{p}_{r}\right)$
can be computed independently by the respective helper, using only
local information, since the quantities $\mathbf{E}_{r}^{H}\left(\mathbf{p}_{r}\right)$
and $\mathbf{g}_{r}\left(\mathbf{p}_{r}\right)$ that appear in $\phi_{r}\left(\mathbf{p}_{r}\right)$
depend solely on the channels from helper $r$ to Bob and Eve, respectively.
This means that \eqref{eq:23} constitutes
a purely decentralized scheme for controlling independently each helper,
leading to the maximization of the system secrecy rate. However, we
should note that since the potentials $\phi_{r}^{o}\left(\mathbf{p}_{r}\right),r\in\mathbb{N}_{R}^{+}$
are generally non convex functions with respect to $\mathbf{p}_{r}$,
the proposed control scheme can only guarantee local optimality and
this mostly depends on the considered approach for wireless channel
modeling.

\section{Experimental Results}

Although the control scheme proposed in Section 4 is only locally
optimal with respect to the positions of the helpers, we present here
numerical simulations, showing that our approach can lead to considerable
savings in terms of both the helper transmitting power and also the
total number of helpers needed for high secrecy rate communication.

\begin{figure}
\centering\includegraphics[scale=0.59]{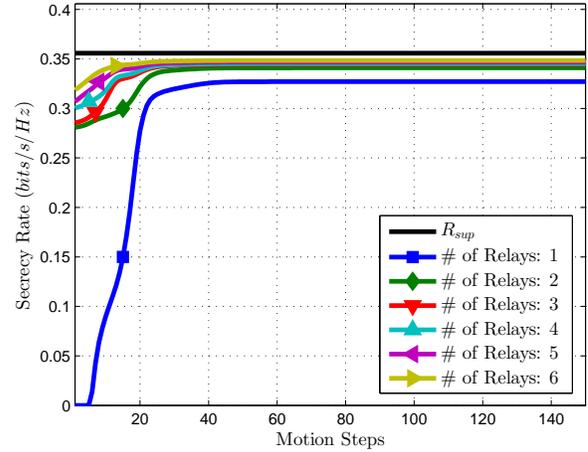}

\caption{\label{fig:helpers}Secrecy rate evolution for various numbers of
helpers.}
\end{figure}

We consider a wireless network in a $\left(3\times5\right)\, m$ rectangular
plane, with of Alice, Bob
and Eve at  points $\left(1.5,0.1\right)$, $\left(1.5,4.9\right)$,
$\left(1.5,4.1\right)$, respectively. We also assume
that the network is assisted by several $2-$antenna helpers performing
cooperative jamming with NN, initially placed with equal spacing across
dimension $x$ and randomly across dimension $y$ with ordinates $2.5+\epsilon$,
where $\epsilon$ is a random variable, with $\epsilon\sim\mathcal{U}\left[-\gamma,\gamma\right]$,
$0<\gamma\ll1$. This choice for the initial arrangement of the helpers
in the plane is justified by the fact that in order to be able to
fairly compare the performance of the system for different numbers
of helpers, the helpers should sense similar channel magnitudes with
respect to Bob and Eve. This is due to path loss. Also, regarding
channel modeling, we assume that $\mu=3.5$, a reasonable value for
mobile robotics applications, and $\lambda=0.4\, m$. 
%This value for $\lambda$ is used in order to facilitate the graphical presentation of our results.
Further, Bob's Signal to Noise Ratio (SNR) is fixed
at $20\, dB$ and the Jamming Noise to Noise Ratio (JNNR) for all
helpers is fixed at $17\, dB$.

Fig. \ref{fig:helpers} shows comparatively the evolution of the secrecy
rate, when the number of the helpers varies from $1$ to $6$, as well
as the upper bound of the system secrecy rate $R_{sup}$, for $150$ motion steps of the proposed
control scheme. In the case where the network is assisted by only
$1$ helper, we observe that whereas the initial value of the secrecy
rate is $0$, after $150$ motion steps, it eventually reaches a value
very close to its supremum. On the other hand, when helper mobility
is not employed, from Fig. \ref{fig:helpers} (rates corresponding to 1 motion step) we can see that in order
for a comparable secrecy rate to be achieved, the number of helpers
must be at least $6$. Therefore, using the proposed
approach, the secrecy rate can be maximized using a much smaller number
of helpers, resulting in large savings in terms of the physical resources
required for a secure, high - secrecy rate system.

Additionally, our approach not only requires fewer helpers, but also
results in smaller helper transmitting power requirements. This fact
can be justified by observing that if, for our example, the total
available number of helpers was constrained to, say, $2$, then, in
order to achieve a secrecy rate close to the $R_{sup}$ without mobility
control, we would have to increase the JNNR (power budgets) of the helpers.

\section{Conclusions}

In this paper, we have addressed the problem of improving the secrecy rate in single-source single-destination networks by employing mobile jammers. We have proposed a novel decentralized motion control scheme for the helpers that effectively maximizes the secrecy rate of the system. Finally, through numerical simulations, we have showed that the proposed scheme can yield considerable savings in terms of both physical and power network resources. This makes our approach very promising for further research and application in cooperative wireless networks.

%Last but not least, Fig. \ref{fig:Cartoon} shows the motion evolution
%(trajectories) of the six helpers considered in this example, with
%respect to the average of Alice's, Bob's and Eve's Rayleigh coefficient
%spatial maps, for intuition. [Dennis, what is the intuition here?]
 
%\begin{figure}
%\centering\includegraphics[scale=0.45]{cartoon}

%\caption{\label{fig:Cartoon}Motion Evolution: Six helpers.}
%\end{figure}

\bibliographystyle{IEEEbib}
%\nocite{*}
\bibliography{OCSRMbib}

\end{document}